\documentclass[usenatbib]{mn2e}

\def\apj{ApJ}%
\def\apss{Ap\&SS}%
\def\aap{A\&A}%
\def\mnras{MNRAS}%
\def\ssr{Space~Sci.~Rev.}%
\title[Efficient planet search]{Efficient analysis in planet transit surveys}

\author[A. Schwarzenberg-Czerny and J.-Ph. Beaulieu]
       {A. Schwarzenberg-Czerny$^{1,2}$ and J.-Ph. Beaulieu$^3$ \\
        $^{1}$ Astronomical Observatory of Adam Mickiewicz University, ul. Sloneczna 36, 60-286 Poznan,
        Poland
       \\
      $^{2}$ Copernicus Astronomical Center, ul. Bartycka 18, PL 00-716 Warsaw,
        e-mail alex{@}camk.edu.pl\\
        $^{3}$ Institut d'Astrophysique de Paris,
UMR7095 CNRS, Universit\'{e} Pierre \& Marie Curie, 98 bis
boulevard Arago, 75014 Paris. France}

\date{Accepted .......
      Received .......;
      in original form ........}

\pagerange{\pageref{firstpage}--\pageref{lastpage}} \pubyear{2004}

\begin{document}

\label{firstpage}

\maketitle


   \begin{abstract}
With the growing number of projects dedicated to the search for
extrasolar planets via transits, there is a need to develop fast,
automatic, robust methods with a statistical background in order
to efficiently do the analysis. We propose a modified analysis of
variance (AoV) test particularly suitable for the detection of
planetary transits in stellar light curves. We show how savings of
labor by a factor of over 10 could be achieved by the careful
organization of computations. Basing on solid analytical
statistical formulation, we discuss performance of our and other
methods for different signal-to-noise and number of observations.
\end{abstract}


\section{Introduction} \label{s1}
The search for extrasolar planets via transits has a venerable
history \citep{stru52}. However, it is the detection of the
transits of HD209458b by \cite{char00}, and the results from the
OGLE survey (Udalski et al., 2002), that have given a very strong
boost to this field, and in the last few years more than 20 ground
based experiments started \citep{hor03}. Although it is very
simple in principle to do a search for extra solar planets via
transits, the small number of positive detections shows that most
of the different projects were over optimistic in their initial
estimates. Independently of the problem of providing reliable
photometry on a large enough number of epochs, problems with the
objects selection and false alarms emerged prominently as shown by
the radial velocity follow up \citep{kon03,bou05}.

Further comments on this issue are found in \citet{alon03} and in
other proceedings of the conferences  ``Scientific Frontiers of
Research on Extrasolar Planets'' \citep{dem03} and ``Extrasolar
planets today and tomorrow'' \citep{bea05}. From a statistical
point of view, the search for transits poses a mere special case
of period search for small signal-to-noise (S/N) ratio with a
known signal shape with short duty cycle spanning over a small
fraction of the phase. On one hand, the space based observations
of COROT \& Kepler of relatively few targets,  are planning the
use of advanced and complex statistical procedures
(\citealt{def01}; \citealt{carp03}; \citealt{jen03} and references
therein). The space surveys differ from the ground
ones discussed here both in terms of the statistics (very long
uninterrupted observations with no atmospheric scintillation) and
different underlying physics (e.g. planetary reflected light with
long duty cycle, c.f. \citealt{jen03}). On the other hand in the
ground surveys data are noisy and interrupted with periodic gaps.
Transits occur with rather short duty cycle and periods comparable
to the gaps period. Several approaches to transit detection in
ground data were already tested in practice, e.g. by \citet{doy00}
and \citet{weld05}. A low success ratio of the ground surveys
calls for massive searches with not much object pre-selection.
This adds motivation for the development of robust and fast new
methods.

The present paper is devoted to modification of the analysis of
variance (AoV) periodogram (Schwarzenberg-Czerny, 1989, Paper I)
for the specific purpose of planetary transit search (AoVtr)
(Sect. \ref{s3}) and to the discussion of related issues of
statistical (Sect. \ref{s3a}) and numerical efficiency (Sect.
\ref{s4}). The properties of AoV related methods known since the
classical work of \citet{fish41} (see also \citealt{fisz63}) are
reviewed in Papers I and by Schwarzenberg-Czerny (1999, Paper II).
The AoV periodo\-gram proved to be an efficient tool in space and
ground based photometric surveys of stellar variability by
Hipparcos \citep{lee97}; OGLE \citep{uda94}; EROS
(\citealt{bea95};\citealt{bea97}). For applications in planet
search see \citep{cum99}.

Detailed analytical results for the AoV transit method are
discussed in Sect. \ref{s3a}. We note that in contrast, no
analytical results are available for other methods. Tingley
(2003a) reviewed  the performances of methods suitable for
planetary transits by resorting to Monte Carlo simulations, and
the result proved to be mixed success. On one
hand, in the original work the results were distorted by
non-optimal implementation of one method. This kind of problem was
difficult to spot because of inherent lack of internal consistency
checks in Monte Carlo simulations. On the other hand,
re-visitation of the problem with the revised implementation
(Tingley, 2003b) yielded the final result constituting a mere
numerical illustration of the general result reached in Paper II.
From a numerical point of view, all the methods discussed by
Tingley (2003a) suffer from a drawback that they demand repeated
calculations for each phase of transit, an increase of workload by
a factor of several dozens. Hence potential advantage of such
phase-independent method as AoVtr introduced here.

\section{On advantage in the periodogram search for variability} \label{s2}

A good comparison of the efficiency of the periodogram and the
period independent variability search methods is illustrated by
the ordinary variance and fast fourier transform (FFT) discrete
power spectrum (DPS), for $N=NF$ observations and frequencies. For
a pure noise input, the suitably normalized power spectrum has
$\chi^2(2)$ distribution with expected value and standard
deviation (s.d.) of $2$ and $\sqrt{2}$, respectively. For the
variance we have $\chi^2(N)$, $N$ and $\sqrt{N}$. Let us add to
the input signal such a sine oscillation that its frequency power
in DPS increases by 2, i.e. by 1.4 s.d. other frequencies
remaining unaffected. By the virtue of Parseval's theorem the
variance is proportional to the sum of PDF, hence its
corresponding increase is by $2/\sqrt{N}$ s.d. For a large $NF=N$
this change in variance becomes entirely insignificant while the
corresponding change at the specific frequency of DPS is
significant. Qualitatively this remains true for comparison of
other frequency independent and dependent searches and for general
uneven sampling.

For existing surveys extending over $NS\sim 10^7$ stars observed
$N\sim10^3$ times over several years, $NF\sim 10^4$ would be
required for proper frequency sampling. For phase folding and
binning the number of operations scales as ${\cal O}(NS\times N\times
NF)\sim 10^{14}$ operations. This circumstance prompted us to
search a way of increasing the efficiency of transit searching methods
in the present article.

At first glance the standard FFT demanding ${\cal O}(NS\times
\log_2{NF}\times NF)$ operations appears to be an attractive
algorithm. Let $NH$ denotes rounded ratio of the orbital period
and transit width. The short duty cycle of transits, of the order
of $1/NH$, where $NH\sim 20$ - $100$ is short. It is well known
that to reach an optimum sensitivity, the resolution of the model
function should match the incoming signal \citep{asc99}. Thus an
efficient detection of a transit with FFT requires $NH=20 - 100$
harmonics, making the total number of frequencies $NH\times NF\sim
10^6$. Then the $\log_2{NF\times NH}$ and $N$ factors become not
so widely different, but the FFT noise would come not from one but
from $NH=20 - 100$ frequencies. However, for contigous data and
smooth light curves of pulsating stars the FFT related algorithms
are methods of choice \citep{pr89}.

\section{Method derivation}\label{s3}

There is no shortage of publications devoted to the
interpretation of phase folded and binned data (see e.g.
\citealt{fisz63} and paper I for the theory and applications,
respectively). The underlying principle is to assume the null
hypothesis, $H_0$, that the data are fitted by a constant value
and then to test it against the alternative hypothesis $H_1(NH)$
employing the phase binned light curve, corresponding to a step function.
Here we adopt only two phase bins of unequal width: in and out of a transit.
Let us assume that width of transit is $v$ in phase units. Then binning corresponds
to fitting the following step function:
 \begin{eqnarray}
s(x)&=&\left\{\begin{array}{cc}a&\mbox{~~~ for } 0\leq x\leq v\\
b&\mbox{~~~ otherwise}\end{array}\right.\label{e3.1}\\
\mbox{where}&~&~\nonumber\\
a&=&<x_{\in T}>\label{e3.2}
\end{eqnarray}
and $N$ and $N_{\in T}$, $<x>$ and $<x_{\in T}>$ denote the number
of observations in total and in the transit and their
corresponding average values. Next we assume that the mean was
subtracted from the data, so that the current mean value vanishes
$0= <x>= N_{\in T}a+(N-N_{\in T})b$, hence
 \begin{eqnarray}
b&=&-\frac{N_{\in T}}{N-N_{\in T}}a\label{e3.3}
\end{eqnarray}

The design of the analysis of variance test for transits (AoVtr)
becomes a special case of \citet{asc89} and \citet{dav90},
in the case where there are only 2 bins. Following
notation of paper II, the AoVtr periodogram statistics $\Theta$ is
defined in terms of sums of squares of model and observations,
denoted respectively $\|x_{\|}\|^2$ and $\|x\|^2$. Normally these
sums are referred as $\chi^2$ statistics. We prefer vector norm
notation where Fisher lemma reduces to Pythagoras theorem. The
corresponding AoV statistics becomes:
\begin{eqnarray}
  \Theta &=& \frac{N-N_{\|}}{1}\frac{\|x_{\|}\|^2}{\|x\|^2-\|x_{\|}\|^2}\mbox{~~~where}\label{e3.4}\\
\|x_{\|}\|^2 &=& N_{\in T}a^2+(N-N_{\in T})b^2 = \frac{N_{\in T}N}{N-N_{\in T}}a^2\label{e3.5}
\end{eqnarray}
and $1$ and $N_{\|}=2$ degrees of freedom account for subtraction
of the average, $<x>=0$, and for one parameter of the model, $a$.
The above procedure easily extends onto weighted observations. In
this case in Eq. (\ref{e3.3}) and (\ref{e3.5}) one should replace
$N$ and $N_{\in T}$ with the corresponding sums of weights.
Elsewhere $N$ remains as the number of degrees of freedom.
Additionally, $<x>$, $<x_{\in T}>$ and $\|x\|^2$ should be
weighted sums.

The implementation of the method is simple. At the beginning we
proceed with binning of observations into $NH$ even phase bins.
Next, we select the bin with the lowest average as transit and
ignore the remaining ones. We exploit here an often forgotten
property, known at least from the times of \citet{whit26}: the
labor at phase folding and binning may be reduced at least by
half, by calculation of bin averages and not their variances. For
the selected bin we calculate $a$ and $\Theta$ from Eq.
(\ref{e3.2}) and (\ref{e3.4}) and use no sum of squares except for
$\|x\|^2$ and $N$ calculated once for all. From the actually
observed value of $\Theta$ and the Fischer-Snedecor $F(,;)$
cumulative distribution $P$ one finds the tail probability
\begin{equation}\label{e3.6}
  Q = NH\left[P\{F(1,N-N_{\|};\Theta)\}-P\{\infty\}\right]
\end{equation}
as an estimates of the detection significance. The $NH$ factor in \ref{e3.6} accounts for the selection of the
transit bin among NH bins in total.

\section{Method performance}\label{s3a}

\subsection{Test power criterion}\label{s3a1}

In order to evaluate the statistical performance of our AoV method
for transits (AoVtr) we apply the test power formalism of paper
II. The higher the test power $1-\beta$, the more sensitive is a
given method, where
  \begin{eqnarray}
~ &~ &R^{-1}(1-\alpha)-R^{-1}(\beta) =
A^2N\frac{\|s_{\|}\|^2}{\sqrt{2N_{\|}}}+ \cdots\label{e3a.1}\\ ~
&~ &\mbox{where}\nonumber\\ ~ &~
&R\left(\frac{\Theta-E}{\sqrt{V}}\right) =
P\{F(1,N-N_{\|};{\Theta})\}\label{e3a.2}
\end{eqnarray}
$1-\alpha$ denotes the significance level, $A^2$ the
signal-to-noise ratio in power units, $N$ and $N_{\|}=2$ are
numbers of observations and parameters of the model, $E$, $V$ and
$P$ denote mean, variance and cumulative distribution of the
Fisher-Snedecor $F(1,N-N_{\|})$ distribution, respectively. For
practical purposes, it is convenient to replace $F$ distribution
in Eq. (\ref{e3a.1}) with the Fisher $Z=(1/2)\log{F}$ distribution
yielding the same information. The latter has near gaussian
distribution (e.g. \citealt{fisz63}).

The actual transit form is assumed to be rectangular of width $v$.
The model consists of two top hat functions of width $c$ and
$1-c$, where $1$ corresponds to the period length. The
corresponding signal shape factor $\|s_{\|}\|^2$ is derived in
Appendix \ref{sa} (Eq. \ref{ea7}). Note, that for fixed
significance level $1-\alpha$, the sensitivity of our method
remains unchanged as long as the expression
$A^2N{\|s_{\|}\|^2}/{\sqrt{2N_{\|}}}$ remains constant.

We adopted a synthetic spectrum of a K0V star ($T_{eff}=5250,
\log{g}=4.5, \log{[M/H]}=0.0, v_{turb}=2 km/s$) from Claret (2000)
and computed synthetic transits over a wide range of filters (from
U to K), and various inclination angles of the system. In this
very broad range, we derived  from Eq. (\ref{ea7})values $1>
\|s_{\|}\|^2 > 0.95$. This demonstrates the small effect of
rectangular approximation on detection efficiency. On the other
hand Eq. (\ref{e87b}) and (\ref{e87d}) demonstrate that using
model transits of width different by factor 2 from the actual one,
yields $\|s_{\|}\|^2\approx 1/2$ causing appreciable loss of the
detection efficiency, corresponding to the use of only half of the
total number of data $N$. At this price one gains factor of
several computation boost by avoiding detailed fit of the transit
width. Our result opens possibility for making informed compromise
between speed and statistical efficiency. Moreover, given the
observing strategy of a given transit survey, it is possible to
estimate in advance the range of periods of transiting planets
that will be probed, and therefore choose an adapted value of
$NH$. Moreover, it is also perfectly possible to do the analysis
in two passes, one with $NH=15$ and a second one with $NH=30$, it
will still be much more efficient than the currently used methods.

We stress that this new scheme is better than the standard binning
scheme adopted for variable star searches based on AoV : in the
particular case of a transit search, we have an a priori knowledge
of the shape of the signal. The signal can be modeled by a top
hat, and  the simpler is the underlying model, in terms of its
parameter count, the more powerful is the test (Eq. \ref{e3a.1}).
The ordinary AoV model has $N_{\|AoV}=NH$ parameters (bin
averages) and the present one, AoVtr, has just $N_{\|AoVtr}=2$
parameters. For matching bin and transit width both models fit the
light curve the same way, $\|s_{\|}\|^2_{AoVtr}=
\|s_{\|}\|^2_{AoV}\approx 1$. Thus for the same signal to noise
ratio $A^2$ the AoV and AoVtr reach similar sensitivity for
detection as long as
$\left.N/\sqrt{N_{\|}}\right|_{AoV}=\left.N/\sqrt{N_{\|}}\right|_{AoVtr}$,
i.e. for $\sqrt{30/2}$ less observations for AoVtr. Further
examples concerning of application of the test power formalism for
design of experiments are provided in Paper II.

\subsection{Comparison with other methods}\label{s3a2}

Paper II employed the test power concept as a
general formalism for evaluation of performance of period search
methods. In particular it demonstrated, that sensitivity for
detection depends on the used signal model and not on the
particular choice of statistics/periodogram. According to paper II
the sensitivity for detection apart from S/N depends on the match
of a light curve and its model implemented in the search method.
However, in this respect any difference between the realistic and
top hat models of transits is small (Sect. \ref{s3a1}). No
difference in performance is expected among different methods
employing the same model for the transit light curve if applied to
the same data. Another fundamental fact in statistics is that
optimum method should involve as few model parameters as
absolutely necessary to decrease residuals (e.g. Paper II). For
this reason sophisticated, multi-parameter models yield poor
sensitivity.

In statistical terms our method is best compared with the matched
filter method (MF) as modified by \citet{ting03b} (mMF). Note that
Eqs. (3) and (4) of \citet{ting03b} are proportional to our Eqs.
(\ref{e3.2}) and (\ref{e3.5}). The difference is we account
properly in Eq. (\ref{e3.4}) for the variance determined from the
same data and yield analytical distribution in return. This should
produce no appreciable difference in statistical performance of
two methods (mMF \& AoVtr). We refer the reader to \citet{ting03a}
and \citet{ting03b} for discussion of mathematical similarity of
the BLS method of \citet{kov02} and MF discussed here.

It remains to demonstrate the relation between the cross
correlation (CCF) and sum of squares ($\chi^2$) statistics:
\begin{equation}\label{e3a.5}
  \chi^2 = \|x-x_{\|}\|^2 = \|x\|^2-2(x,x_{\|})+\|x_{\|}\|^2
\end{equation}
The last term above is non-random (a constant), the first one is
sum of many terms, therefore random with small variation. Both are
independent of frequency. The term
with the dominant variance is the middle one as it reduces to the
sum of few terms in transit. This term corresponds to the CCF. So,
except for sign and (nearly) constant shift the distributions of
$\chi^2$ and CCF are identical and yield identical statistical
conclusions. This applies in general to the MF approach as pursued
by \citet{weld05} and \citet{jen03}. Note that in the latter
case the employed model, $x_{\|}$ is different from the transit
one. The best known CCF-like statistics is power spectrum,
consisting of sum of squared norms of sine and cosine CCF
functions. In time series context \citet{lom76} first demonstrated
for power spectrum statistical equivalence of CCF and $\chi^2$
statistics.

Comparison with \citet{doy00} who employed the MF method with the
sum of absolute value of residuals as the test statistic is more
difficult. For gaussian errors the quadratic norm used in AoVtr
and its CCF equivalent in some MF methods has optimum properties
(e.g. \citealt{ead71}). However, for certain distributions with
large or no moments the absolute value statistics is known to
perform better than the quadratic one. In such applications
\citet{doy00} method may be better than those discussed so far.

On one hand performance of the bayesian method of \citet{def01}
in tests by \citet{ting03b} was poor. On the other hand
Schwarzenberg-Czerny (1998) invoked Wold theorem to demonstrate
that in the asymptotic limit of large data set performance of
bayesian methods should match that of the classical ones, for
similar setup. This result is also in consistency with
\citet{gre92}. In this respect poor performance could arise e.g.
from implementation inconsistency and/or from small data behaviour.

\subsection{Realistic application}\label{s3a3}

We applied the AoVtr method to the publicly available 142 OGLE
light curves of periodic transit candidates (Udalski et al.,
2002). With $NH=30$ for 139 light curves we detect the signal with
the same principal period always with $\Theta>15$. For the
remaining 3 the period claimed originally appeared as an alias.

\section{Efficient algorithm implementation}\label{s4}

For as large gaps as encountered in astronomical observations from
the ground the FFT related methods discussed in Sect. \ref{s2}
suffer from large overhead for processing null data in the gaps.
The fastest known methods for observations with large gaps rely on
phase folding and binning of data as in the case of AoV and AoVtr
methods. In the simplest implementation for each frequency one
calculates phases of observations and assigns into respective
phase bins. For each observation falling in a given bin, the
weight and weighted sum of this bin are incremented. This is the
most labor consuming part. A known drawback of the phase binning
is the loss of the detection efficiency for the eclipses/transits
falling at the bin boundary \citep{asc99}. An efficient protection
against that is to bin the observations starting at different
initial phases \citep{ste78}. Each such bin set is called a
coverage. The trick preventing repeated summing of observations
for each coverage is to first bin the observations into sub-bins.
Than the whole-bin sums for each coverage are obtained at the end
by summing the corresponding sub-bins, with negligible overhead.

Simple fixed bin size implementation of phase folding is prone to
occasional failure, both in the numerical and statistical sense,
due to incomplete phase coverage. But we found that a robust yet
statistically correct solution for the treatment of these seldom
occurring cases is to sort observations in phase and then to bin
them evenly according to their sequence number. Now, provided that
for all phases equal 0 the sort routine preserves the original
order intact, our code would work also for 0 frequency. The
periodogram value at 0 frequency is nothing else but a frequency
independent AoV variability test.

Our sample code implementation in {\tt C} taking care of both procedures
is presented in Table \ref{t1}. We stress that the {\tt floor} operation should
be implemented by simple register shifts, back and forth, with
null filling. Then in the innermost loop there remain only 4 other
floating point operations per star, observation and
frequency. For two coverages this yields a factor over 4 labor
saving compared to the calculation of variances for each bin and
for each coverage in separate.

Note, that the same sub-binning concept enables the implementation
of the AoVtr method with variable width of transits. This requires
an extra piece of code checking whether inclusion of additional
neighbor sub-bins increases or decreases detection significance.
To facilitate that, one should store a pre-computed table of
critical F values for $(1,n), n=1,N$ degrees of freedom. Note that
values for large $n$ are going to be rarely used and may be
omitted.

\section{Optimum frequency sampling}\label{s5}

The proper selection of the frequencies to be scanned poses a challenge for efficient design
of the analysis. Nominally adapted sampling in period search
corresponds to such a step in frequency $\delta\nu_p$, that over
the whole interval of observation $\Delta t_p$ any essential light
curve feature remains marginally in phase. For a sine curve that
corresponds to $\Delta t_p\delta\nu_p\leq 1$. However, for
transits and other short duty cycle processes the condition
becomes even more stringent $\Delta t_p\delta\nu_p\leq 1/NH$, and
therefore the sampling in frequency would have to be even more dense.

Instead we propose a two-tier approach bound at saving a factor of
several computing effort. The aim of the first scan of data is to
detect with high sensitivity any periodic variability, with no
guarantee of proper period identification. At this stage we take
full advantage of otherwise annoying presence of aliases. Thus we
strive at detection of any of the aliases as an indication of the
presence of a periodic signal. Due to their even spacing, the
detection of aliases remains efficient even for severe
undersampling. This occurs due to a vernier effect as described by
\citet{asc05}. Except for the pathological case of
commensurability of steps, one of loosely and evenly spaced
sampling frequencies ought to coincide with one of the evenly
spaced aliases. This ensures detection but yields no reliable
period value. Yet at this stage are rejected all constant stars,
reducing our sample by a factor up to 100. At the next stage it
remains to recalculate with proper sampling the periodogram for
the variable stars detected at stage one. The difficult task of
selecting among several aliases remains, but one gets to that
point by a less exhausting way.

To explain the role of aliases and the vernier principle let us
remind of simple facts about effect of sampling on Fourier
transform and its norm, DPS. The final discrete sampling pattern
$w$ may be approximated by product of three functions $w = d\times
s\times p$ representing a discrete pattern of individual exposures,
seasonal pattern and the total duration of the observing campaign. The
corresponding time scales $\delta t_d<\Delta t_s<\Delta t_p$, are
minutes to days, several month and several years, respectively. In
the frequency space they correspond to the total Nyquist range.
This range is covered by evenly spaced alias peaks for any single real
period and width of an individual peak, $\Delta \nu_d>\delta
\nu_s>\delta \nu_p$, where $\Delta \nu_d=1/\delta t_d$ and so on
for $s$ and $p$. For simplicity we concentrate here on the most
relevant alias pattern due to seasons. Each consecutive sampling
point becomes shifted by a constant step with respect to the
aliases.  The case of commensurable steps of aliases and sampling constitutes a rare
pathology. Since aliases are separated by troughs, there are
about $\delta \nu_s/2\delta \nu_p=\Delta t_p/2\Delta t_s$ aliases
associated with each single oscillation. In order to detect any of
them, suffices if say 3 or 5 of the sampling frequencies cover the
whole pattern of width $\delta \nu_s$, i.e. if the periodogram is
sampled with the step $\delta\nu_v=1/4\delta t_s$. This
constitutes a gain by a factor $\delta\nu_v/\delta\nu_p=\Delta
t_p/4\Delta t_s$ corresponding numerically to duration of a
project, in units of a year.

In practical terms the suitable frequency step $\delta\nu_v$ is
best selected by trial and error on a survey sub-sample. A
suitable undersampling step choice should not produce any large
loss of detections compared to the case of proper sampling by
$\delta\nu_p$.

\section{Non-gaussian errors}\label{s5a}
The referee raised important issue of possible
non-gaussian distribution of errors (the parent distribution).
There is no space here for a thorough discussion of this problem,
but some points are worth of mention. Applicability of the F
statistics in the AoV method depends on the sum of squares in the
numerator and denominator of Eq. (\ref{e3.4}) obeying the $\chi^2$
distribution. In this respect more critical is the numerator
$\|x_{\|}\|^2$ (Eq. \ref{e3.5}). For gaussian parent distribution
$\|x_{\|}\|^2$ obeys the $\chi^2(1)$ distribution by virtue of the
Fisher lemma, where $1=N_{\|}-1$. If non-gaussian errors satisfy
assumptions of the Central Limit Theorem, then in the asymptotic
limit of a large number of transit observations, $N_{\in
T}\rightarrow\infty$, the average $<x_{\in T}>$ obeys the gaussian
distribution and its square in Eq. (\ref{e3.5}) obeys $\chi^2(1)$,
as required. The relevant assumption of the Central Limit Theorem
is existence of bounds on the moments of the parent distribution.
This is not as restrictive as it may appear. Usually observations
are pre-screened so that their errors have limited magnitude hence
all moments of the distribution exist.

The real issue is whether the number of observations per bin, i.e.
here per transit, $N_{\in T}$, is sufficient to approach the
asymptotic limit. From the proof of the Central Limit Theorem
follows that the relevant merit figures for symmetric and
asymmetric parent distributions are $\mu_3/(3! \sqrt{N_{\in T}})$
and $\mu_4/(4! N_{\in T})$, respectively, where $\mu_i$ denotes
$i$-th central moment of the parent distribution in units of
$i$-th power of its standard deviation $\sigma$ (e.g.
\citealt{bran70}). Except for pathological parent distributions with
large high moments, $\mu_i\gg\sigma^i$, smallness of the merit
figures indicates approaching of the asymptotic limit. Let us
consider a particular case of strongly asymmetric parent
distribution $\chi^2(2)$, i.e. the $e^{-x}$ distribution with
$\mu_i\sim\sigma^i$. The average $<x_{\in T}>$ obeys the
$\chi^2(2N_{\in T})$ distribution. \citet{fish25} argued
that for $N_{\in T}>30$ the asymptotic limit is good enough. His
conclusion is subject to the restriction that the significance
level does not exceed the usual range of $0.999$. In order to get
$N_{\in T}=30$ observations in transits, one needs $N \sim 30 NH =
1000$ observations in total, on average. Thus our analytical
theory should apply directly for a number of existing transit
surveys with $N>1000$ observations per candidate object.

\section{Conclusions}\label{s6}

We presented arguments for the adoption of a new AoV related test
particularly suitable for detection of planetary transits in
stellar light curves. As it is based on just one parameter fit its
statistical test power is bound to exceed that of common
variability tests. We demonstrated how by careful organization of
computations savings of labor by a factor of over 10 may be achieved.

\section*{Acknowledgments}
Micha{\l} Szyma\'{n}ski is thanked for asking questions, for which
we found answers only now. Present collaboration was made
possible due to support in part by the  Polish-French  Associated European Laboratory
LEA Astro-PF. ASC acknowledges support by KBN grant 1P03D 025 29.

\bsp

\appendix

\section{Projection of signal onto model space}\label{sa}

For the test signal we shall use the function $s$ from Eq.
(\ref{e3.1}) with constants $a$ and $b$ redefined so that
$<s>=(s,1)=0$ and $\|s\|^2=(s,s)=1$. However, in the present
consideration scalar products involving sums over observations
should be replaced with integrals of function product over the
entire range of phases:
$(f,g)=\int_0^1f(\varphi)g(\varphi)d\varphi$. In such a case we
obtain
 \begin{eqnarray}
a&=&\sqrt{\frac{v}{1-v}}\label{ea5}\\b&=&-\frac{1-v}{v}a\label{ea6}
\end{eqnarray}

The norm of the signal $s$ projected onto its model function,
$\|s_{\|}\|^2$ has to be calculated following the prescription from paper II:
  \begin{equation}
\|s_{\|}\|^2=\sum_{l=1}^2|<s,\phi^{(l)}>|^2\label{ea1}
\end{equation}
We assume that the test signal and transit model are rectangular of different width,
$v$ and $c$, respectively. The orthonormal model functions
functions $\phi^{(l)}(x)$ for AoVtr, corresponding to in/out transit phases are
adopted as:
 \begin{eqnarray}
\phi^{(1)}(x)&=&\left\{\begin{array}{cc}\frac{1}{\sqrt{c}}&\mbox{~~~ for } 0\leq x\leq c\\
0&\mbox{~~~ otherwise}\end{array}\right.\label{ea2}\\
\phi^{(2)}(x)&=&\left\{\begin{array}{cc}\frac{1}{\sqrt{1-c}}&\mbox{~~~ for } c\leq x\leq 1\\
0&\mbox{~~~ otherwise}\end{array}\right.\label{ea3}\\
\end{eqnarray}
Substituting these definitions into Eq. (\ref{ea1}), for $v\geq c$ one obtains
  \begin{equation}
\|s_{\|}\|^2=\frac{(1-v)c}{(1-c)v}\label{ea7}
\end{equation}
For $v<c$ one should swap $v$ and $c$ in the above equation.
The following particular results are of interest here:
 \begin{eqnarray}
\|s_{\|}\|^2&\leq& 1\label{e87a}\\
\|s_{\|}\|^2&\rightarrow&\min(\frac{c}{v},\frac{v}{c}) \mbox{~~for~~}v,c\rightarrow 0\label{e87b}\\
\|s_{\|}\|^2&\rightarrow& 1 \mbox{~~~ for~~~}v\rightarrow c\label{e87c}\\
\|s_{\|}\|^2&=&\frac{NH-2}{2(NH-1)} \mbox{~~ for~~}v=2/NH, \;c=1/NH\label{e87d}
\end{eqnarray}

\section{List of symbols}\label{sl}

\begin{description}
\item[$A$] - signal-to-noise amplitude ratio;
\item[$a$] - parameter of the model (brightness in transit);
\item[$b$] - parameter of the model (brightness out of transit);
\item[$c$] - transit width in the model signal, $c=1/NH$;
\item[$E$] - the expected value of a distribution;
\item[$FFT$] - fast fourier transform;
\item[$F(N_1,N_2;\cdot)$] - the Fisher-Snedecor probability density distribution for $N_1$ and $N_2$ degrees of
freedom, abbreviated as $F(N_1,N_2)$;
\item[$H_0$] - the statistical null hypothesis stating that a signal consists of
pure noise;
\item[$H_0(NH)$] - the statistical alternative hypothesis stating that a signal consists of
noise plus deterministic component (e.g. transit of width $1/NH$);
\item[$i=p,s,d,v$ sampling indices] - correspond to (p) proper
sampling/full span of observations, (s) sesonal span/frequency
pattern (1 yr), (d) day span/frequency pattern and (v) the vernier
sampling pattern;
\item[$N$] - total number of observations;
\item[$NF$] - number of frequencies in the periodogram;
\item[$NH$] - Period length in units of transit width, also optimum number of bins
for ordinary no top-hat AoV method;
\item[$NS$] - number of stars observed in a whole survey;
\item[$N_{\in T}$] - number of observations in transit;
\item[$N_{\|}\equiv N_{\|\;AoVtr}$] - number of parameters of the top-hat model,
$N_{\|}=2$;
\item[$N_{\| AoV}$] - number of parameters of the phase binned;
model $N_{\| AoV}=NH$;
\item[$P$] - the cumulative probability distribution;
\item[$R(0,1;\cdot)$] - a normalized probability distribution,
such that $E\{R\}=0$ and $V\{R\}=1$, e.g. normalized $F$,
normalized $Z=(1/2)\log{F}$ or normal distribution;
\item[$S/N$] - signal-to-noise power ratio, i.e. ratio of
squared amplitudes $S/N=A^2$;
\item[$s$] - the normalized actual deterministic
signal, $<s>=0$ and $\|s\|^2=1$;
\item[$\|s_{\|}\|^2$] - squared normalized projection of the actual deterministic
signal $s$ onto the assumed top hat model, $x_{\|}$, corresponding
to the squared cosine between vectors $s$ and $x_{\|}$, hence
$\|s_{\|}\|^2 = (s,x_{\|})^2 / \{\|s\|^2 \|x_{\|}\|^2\}$;
\item[$T$] - integration variable for $\Theta$ F statistics;
\item[$V$] - the variance value of a distribution;
\item[$v$] - transit width in the actual signal;
\item[$x$] - values of observations, $<x>=0$;
\item[$x_{\|}$] - values of model light curve, $<x_{\|}>=0$, in
terms of the orthogonal components
$x_{\|}(t)=\sum_{l=1}^2(x_{\|},\phi^{(l)})\phi^{(l)}(t)$;
\item[$x_{\in T}$] - values of observations in transit, $<x_{\in
T}>=a$;
\item[$\delta\nu_i$] - where $i=p,s,d,v$, the frequency i.e. periodogram sampling
interval, $\delta\nu_i=1/\Delta t_i$;
\item[$\Delta\nu_i$] - where $i=p,s,d,v$, the total frequency span
of a feature or of the periodogram;
\item[$\delta t_i$] - where $i=p,s,d,v$, the time i.e. observation sampling
interval;
\item[$\Delta\nu_i$] - where $i=p,s,d,v$, the time span of a feature or the
observations;
\item[$\phi^{(l)}(t)$] - normalized top hat functions covering
transit $(l=1)$ and out of transit $(l=2)$ bins;
\item[$\chi^2(M;\cdot)$] - the $\chi^2$ probability density distribution for $M$ degrees
of freedom, abbreviated as $\chi^2(M)$;
\item[$\Theta$] - the observed value of F statistics;
\item[$(\cdot,\cdot)$] - scalar product (possibly weighted);
\item[$<\cdot>$] - average value of argument, $<x>\equiv(1,x)$;
\item[$\|\cdot\|^2$] - quadratic norm of argument, i.e. (possibly
weighted) sum of squares $\|x\|^2\equiv(x,x)$;
\end{description}

\section{Source code}\label{sb}
Sample implementation of the transit periodogram in C code is
presented in Table \ref{t1}.
\setcounter{section}{3}
\begin{table*}
\begin{minipage}{180mm}
\caption{\normalsize Sample C implementation of the transit
periodogram.\label{t1}}
 \framebox[180mm][l]{\begin{tabular}{ll}\begin{minipage}{83mm}\small\begin{flushleft}
int aov (int nobs, TIME tin[], FLOAT fin[], int nh, int ncov,\\
~~~~~LONG nfr, TIME fr0, TIME frs, FLOAT * th)\\
\{ \\ /* (C) by Alex Schwarzenberg-Czerny, 2003, 2005 */ \\
  int nct[MAXBIN],ind[MAXOBS], i, ibin, ip, nbc;\\
  LONG ifr, iflex;\\
  FLOAT f[MAXOBS], ph[MAXOBS], ave[MAXBIN], af, vf, sav;\\
  TIME t[MAXOBS], fr, at, dbc, dph;\\~\\
  if (((nh+1)*ncov $>$ MAXBIN) $||$ (nobs $>$ MAXOBS) $||$ \\~~~~ (nobs $<=$ nh+nh))\\
    \{\\
/*~~    fprintf(stderr,"AOV: error: wrong size of arrays/n"); */ \\
~~  return(-1); \\
    \};\\
  nbc = nh * ncov;\\
  dbc = (TIME) nbc;\\
/*  calculate totals and normalize variables */ \\
  iflex = 0; at = (TIME) (af = vf = (FLOAT)0.); \\
  for (i = 0; i $<$ nobs; i++) \{ af += fin[i]; at += tin[i]; \} \\
  af /= (FLOAT) nobs; at /= (TIME) nobs; \\
  for (i = 0; i $<$ nobs; i++) \\
    \{ \\
     ~~ t[i] = tin[i] - at; \\
     ~~ f[i] = (sav = fin[i] - af); \\
     ~~ vf += sav*sav; \\
    \}; \\
 /* assumed: sum(f[])=0, sum(f[]*f[])=vf and sum(t[]) is small
 */\\   for (ifr = 0; ifr $<$
nfr; ifr++) /*  Loop over frequencies
  */\\
    \{ \\
~~fr = ((TIME) ifr) * frs + fr0;\\ ~~for (ip = 0; ip $<$ 2 ; ip++)
\\ ~~\{\\
      ~~~~for (i = 0; i$<$ nbc; i++) \{ ave[i] = 0.; nct[i] = 0; \};\\
      ~~~~if ( ip == 0) /* Try default fixed bins ... */\\
\framebox[83mm][l]{\begin{minipage}{83mm}\small\begin{flushleft}
      ~~~~~~for (i = 0; i $<$ nobs; i++) /* MOST LABOR HERE */\\
        ~~~~~~\{\\
            ~~~~~~~~dph=t[i]*fr; /* TIME dph, t, fr
            */\\
            ~~~~~~~~ph[i]=(sav=(FLOAT)(dph-floor(dph)));\\
            ~~~~~~~~ibin=(int)floor(sav*dbc);\\
            ~~~~~~~~ave[ibin] += f[i];\\
            ~~~~~~~~++nct[ibin];\\
        ~~~~~~\}  \end{flushleft}  \end{minipage}}~\\
 \end{flushleft}  \end{minipage} &
 \begin{minipage}{85mm}\small\begin{flushleft}
      ~~~~else /* ... and elastic bins, if necesseary */\\
      ~~~~\{  \\
        ~~~~~~++iflex;            /* sort index ind using key ph */\\
        ~~~~~~sortx(nobs,ph,ind); /* corrected NR indexx would do */\\
        ~~~~~~for (i = 0; i $<$ nobs; i++)\\
        ~~~~~~\{\\
            ~~~~~~~~ibin=i*nbc/nobs;\\
            ~~~~~~~~ave[ibin] += f[ind[i]];\\
            ~~~~~~~~++nct[ibin];\\
        ~~~~~~\};\\
      ~~~~\};\\
  /* counts: sub-bins$=>$bins */\\
      ~~~~for (i=0;i$<$ncov;i++) nct[i+nbc]=nct[i];\\
      ~~~~ibin=0;\\
      ~~~~for(i=ncov+nbc-1;i$>$=0;i--) nct[i]=(ibin+=nct[i]);\\
      ~~~~for (i=0;i$<$nbc;i++) nct[i]-=nct[i+ncov];\\
      ~~~~for (i = 0; i $<$ nbc ; i++) /* check bin occupation */\\
        ~~~~~~if (nct[i] $<$ CTMIN) break;\\
      ~~~~if (i$>$=nbc) break;\\
    ~~\};\\
    ~~\\
~~/* data: sub-bins$=>$bins */\\
    ~~for (i=0; i$<$ncov; i++) ave[i+nbc]=ave[i]; \\
    ~~sav=0.;\\
    ~~for (i=ncov+nbc-1;i$>$=0;i--) ave[i]=(sav+=ave[i]);\\
    ~~for (i=0;i$<$nbc;i++) ave[i]-=ave[i+ncov];\\
    ~~\\
      \framebox[78mm][l]{\begin{minipage}{78mm}\small\begin{flushleft}
~~/* AoV statistics for transits */\\
    ~~sav=ave[0]/nct[0];\\
    ~~for (i=0;i$<$nbc;i++) \\
    ~~~~if((ave[i]/=nct[i])$>=$sav) \{sav=ave[i];ibin=i;\};\\
    ~~sav*=sav*nct[ibin]*nobs/(FLOAT)(nobs-nct[ibin]); \\
    ~~th[ifr] = sav/MAX(vf-sav,1e-32)*(nobs-2);\\
  \}; /* where 'vf' keeps the total sum of squares of f[]*/\end{flushleft}  \end{minipage}}~\\
~~/* the same for the ordinary AoV statistics: \\
    ~~sav=0.;\\
    ~~for (i=0;i$<$nbc;i++)  sav+=(ave[i]*ave[i]/nct[i]);\\
    ~~sav/=(FLOAT)ncov; \\
    ~~th[ifr] = sav/(nh-1)/MAX(vf-sav,1e-32)*(nobs-nh);\\
    \}; */\\
/* if (iflex $>$ 0) fprintf(stderr, "aov:warning:" \\
~~~~ "poor phase coverage at \%d frequencies/n",iflex); */ \\
  return(0);\\
 \};
 \end{flushleft}  \end{minipage}
\end{tabular}}
{\small\vspace{1.5mm}\newline\begin{tabular}{ll}\begin{minipage}{83mm}
Input:\begin{description}\item{\bf nobs -} number of
observations;\item{\bf tin[nobs], fin[nobs] -} times and values of
observations;
\item{\bf nh, ncov -} number of phase bins and number of
coverages;
\item{\bf nfr - } number of frequencies;
\item{\bf fr0, frs -} frequency start and step;
\end{description}
  \end{minipage} &
 \begin{minipage}{85mm}
Parameter:\begin{description}
\item{\bf TIME -} extended precision type,\newline
e.g. \#define TIME double
\item{\bf MAXBIN -} maximum (nh+1)*ncov;
\item{\bf MAXOBS -} maximum number of observations;
\item{\bf CTMIN -} minimum bin occupancy$>$1\newline
e.g. \#define CTMIN 5
\end{description}

Output:\begin{description}
\item{\bf th[nfr] -} the AoV periodogram;
\end{description}
\end{minipage}
\end{tabular}}\vspace{1.5mm}
{\raggedright\small\em The complete source code
and a test example may be downloaded from the web address \verb+http://www.camk.edu.pl/~alex/#software+}
\end{minipage}
\end{table*}\label{lastpage}
\end{document}